\documentclass[aps,prb,reprint,showkeys]{revtex4-1}
\usepackage{bm,graphicx,tabularx,array,dcolumn,xcolor,microtype,multirow,amscd,amsmath,amssymb,amsfonts,physics,diagbox}
\usepackage[version=4]{mhchem}
\usepackage[utf8]{inputenc}
\usepackage[T1]{fontenc}
\usepackage{txfonts}
\usepackage{array}
\usepackage{makecell}

\usepackage{dblfloatfix,placeins,float,physics}

\usepackage[
	colorlinks=true,
    citecolor=blue,
    linkcolor=blue,
    filecolor=blue,      
    urlcolor=blue,	
    breaklinks=true
	]{hyperref}
\usepackage{caption} 
\usepackage{subcaption}

\begin{document}

\title{The role of steric effects on hydrogen atom transfer reactions}
\author{Yi Sun, Jacob N. Sanders, K. N. Houk}
\email{houk@chem.ucla.edu}
\affiliation{Department of Chemistry and Biochemistry, University of California, Los Angeles, California 90095, United States}

\date{\today}

\begin{abstract}
We explored how steric effects influence the rate of hydrogen atom transfer (HAT) reactions between oxyradicals and alkanes. Quantum chemical computations of transition states show that activation barriers and reaction enthalpies are both influenced by bulky substituents on the radical, but less so by substituents on the alkane. The activation barriers correlate with reaction enthalpies via the Evans-Polanyi relationship, even when steric effects are important.  Dispersion effects can additionally stabilize transition states in some cases.
\end{abstract}

\maketitle
\raggedbottom

\section{Introduction}
Hydrogen atom transfer (HAT) reactions are one of the most fundamental chemical reactions:

\begin{figure}[H]
\includegraphics[width=\linewidth]{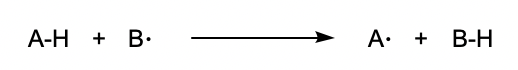}
\caption{The hydrogen atom transfer reaction.}
\centering
\end{figure}

These reactions are key steps in a wide range of chemical, environmental, and biological processes. Recently, a number of photoinduced \cite{Zhang, Deng, Cao} and electrochemical \cite{Niko} HAT reactions that have value in synthetic organic chemistry have been reported. There are also numerous case studies of HAT reactions that involve metal species \cite{Joseph, Zass, Cheong, Laxmi}.

There have been many studies aimed at understanding the barrier heights of HAT reactions. A celebrated example is the Bell-Evans-Polanyi principle, \cite{Bell,EP} which can be expressed mathematically as
\begin{equation} \label{eq:1b}
\Delta H^{\ddag} = E_0 + \alpha \Delta H
\end{equation}
in which $\Delta H^{\ddag}$ is the activation barrier of the reaction, i.e., barrier height, $\Delta H$ is the enthalpy change of the reaction, and $E_0$ and $\alpha$ are empirical fitting parameters. In particular, the parameter $\alpha$ is related to the position of the transition state along the reaction coordinate.

A refinement of the Bell-Evans-Polanyi principle is the Roberts-Steel relationship, which also accounts for polar and conjugation effects \cite{RS}. Recent work by Liu \emph{et al.} \cite{Liu} has proposed a simplified Roberts-Steel relationship in which the activation barrier of the reaction,$\Delta H^{\ddag}$, depends on two quantities: the reaction enthalpy $\Delta H$ and the difference in the electronegativity $\Delta \chi$ of the radicals. This gives rise to the equation

\begin{equation} \label{eq:2}
\Delta H^{\ddag} = \alpha \Delta H + \beta \Delta \chi^2 + \gamma
\end{equation}
in which $\alpha$, $\beta$ and $\gamma$ are empirical fitting parameters. 

Another strategy to understand the barriers of HAT reactions is to use the Marcus cross relationship, which was shown by Mayer \cite{Mayer} to relate the HAT reaction barrier heights to the self-exchange rate constants and the chemical equilibrium  constant of the HAT reaction via the equation below:

\begin{equation} \label{eq:1a}
k_{AH/B} = \sqrt{k_{AH/A}k_{BH/B}K_{eq}f}
\end{equation}

where $k_{AH/A}$ and $k_{BH/B}$ are the rate constants for the respective self-exchange HAT reactions, and $K_{eq}$ is the equilibrium constant of the HAT reaction, which can be calculated from the change in Gibbs free energy, $\Delta G$, of the reaction; $f$ is a dimensionless proportionality factor. Thus, Marcus theory relates HAT reaction barriers to kinetic factors, namely the intrinsic barriers of self-exchange, and to thermodynamic factors, namely the free energy change of the HAT reaction.   By contrast, the modified Roberts-Steel relationship uses intrinsic properties of reactant and product radicals, namely their electronegativities, along with thermodynamic factors, namely the enthalpy of reaction, to understand reaction barriers.

More recently, the Hong group has developed machine learning models to correlate rates of HAT reactions with many different properties of the reactants and products.\cite{Hong}  We have been carrying out parallel efforts to understand the fundamental factors controlling rates.  In this paper, we explore the role, if any, of steric effects on the rates of HAT reactions.

\section{STERIC EFFECTS IN ORGANIC CHEMISTRY}

Although the relationships discussed in the previous section have been used to explore many types of HAT reactions, these studies have included few sterically hindered substrates or radicals.  Accordingly, we sought to determine whether steric effects are influential in HAT reactions.  On the one hand, bulky groups might slow HAT by increasing the energies of transition states via steric repulsion as atoms attempt to approach each other more closely than the sum of their van der Waals radii \cite{Bondi}.  On the other hand, if atoms just touch at the sum of their van der Waals radii in the transition states, then dispersion interactions may actually accelerate the HAT reaction \cite{Peter}.  Such dispersion effects have been noted in a number of hemolysis and radical recombination reactions. \cite{Peter}

Steric effects are well known throughout organic chemistry. When atoms come too close, such that the distance between them is shorter than the sum of their van der Waals radii, the energy  rises \cite{Bondi}; in this way steric hindrance can slow down or even prevent a reaction. More recently, the importance of steric attraction, which arises from London dispersion forces between large substituents, has been emphasized \cite{Peter}. There are many examples where steric effects influence the reactivity of molecules and the selectivity of substitution reactions \cite{Tolman, Crossley, Bo, Zhou}; they can even determine whether certain radicals are stable in the first place \cite{Hicks, Power}. Nevertheless, systematic studies of steric effects in HAT reactions across a wide range of substrates are scarce. 

We have investigated the role of steric effects in HAT reactions between alkoxy and phenoxy radicals and alkane substrates. We observe the largest steric effects for alkoxy and phenoxy radicals and only smaller steric effects among substrates.  In particular, steric attraction is often important in controlling barrier heights. Remarkably, we discover that steric effects influence not only barrier heights but also reaction enthalpies. Because steric effects influence reaction enthalpies, we find that that activation barriers remain well-connected to reaction enthalpies via the Evans-Polanyi relationship even in the presence of bulky substituents.

\section{Computational method}

Ground and transition structures were optimized at the $\omega$B97xD/6-31G(d) level of theory \cite{Francl, Harihanan, Chai, Chai2} using Gaussian09 \cite{Gaussian}. Optimized structures were confirmed via frequency calculations, with ground states having all positive frequencies and transition states having one imaginary frequency corresponding to the reaction coordinate.  These frequencies were also used to compute enthalpies and Gibbs free energies at 298 K and a standard state of 1 atm.  Energies were corrected via single-point computations performed at $\omega$B97xD/6-311G++(d,p) level of theory.

\section{The investigated reactions and the results}

We explored 120 HAT reactions between 10 radicals and 12 alkanes. The structures of the alkoxy, phenoxy, and cyclohexoxy radicals are shown in Figure 2, while the structures of alkane substrates are shown in Figure 3. All HAT reactions studied are between oxy radicals in Figure 2 and the illustrated C–H bonds of the alkanes in Figure 3. To minimize polar effects, no polar substituents or heteroatoms were studied. All barrier heights are shown in Table I, with the corresponding reaction enthalpies shown in Table II.

\begin{figure}[H]
\includegraphics[width=\linewidth]{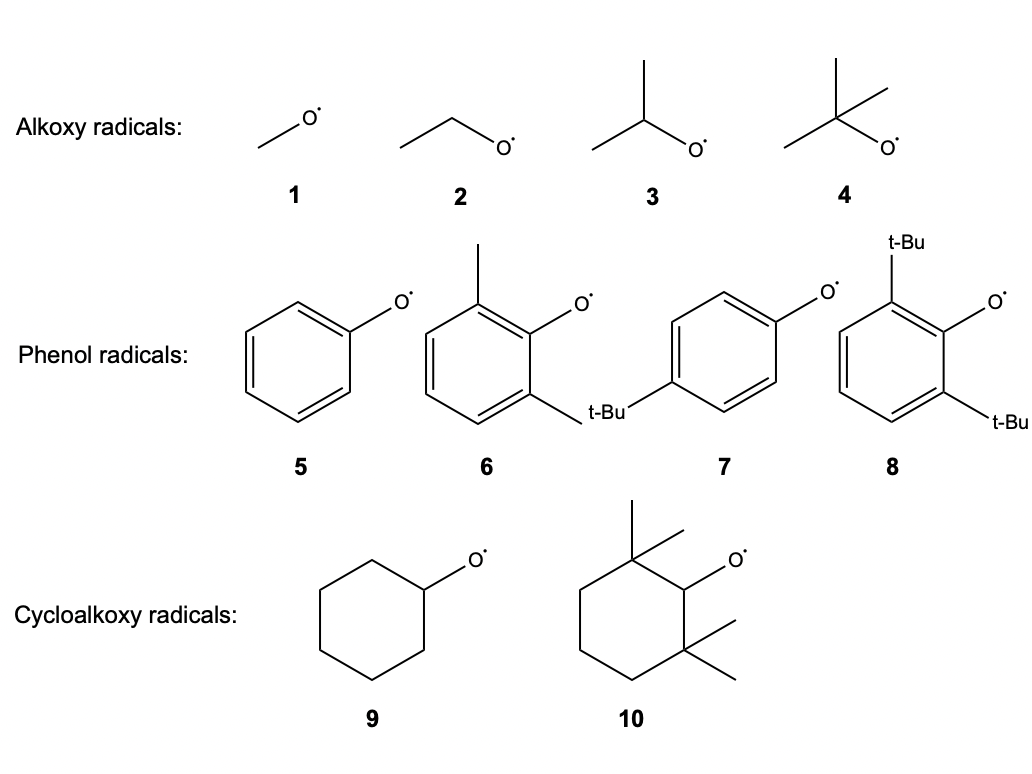}
\caption{The 10 radicals studied in this work.}
\centering
\end{figure}

\begin{figure}[H]
\includegraphics[width=0.85\linewidth]{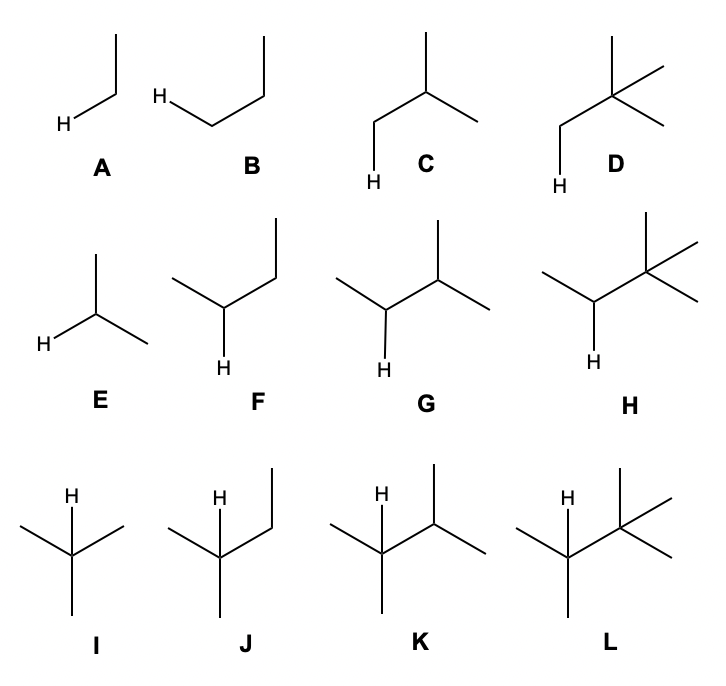}
\caption{The 12 alkanes studied in this work with the abstracted hydrogen drawn explicitly.}
\centering
\end{figure}

\begin{table}[H]
\centering
\begin{tabular}{|c|c|c|c|c|c|c|c|c|c|c|}
\hline
\backslashbox{Alkane}{Radical}   & 1   & 2   & 3   & 4    & 5    & 6    & 7    & 8    & 9    & 10   \\ \hline
A & 6.9 & 7.7 & 6.1  &  5.1  & 19.5 & 21.5 & 20.6 & 26.3 & 5.7  & 8.7  \\ \hline
B & 7.1 & 8.2 & 6.6 &  5.4  & 19.9 & 22.0 & 21.4 & 26.1 & 6.0  & 8.8  \\ \hline
C & 6.5 & 7.6 & 5.9 & 5.0  & 19.3 & 23.0 & 20.6 & 25.5 & 5.3  & 8.1 \\ \hline
D & 6.1 & 7.1  &  5.3  & 4.4  & 19.9 & 22.7 & 21.0 & 25.5 & 5.0  & 8.0  \\ \hline
E & 4.1 & 4.6 & 3.0 & 2.7 & 15.7 & 18.4 & 17.1 & 22.0 & 2.7  & 5.4  \\ \hline
F & 4.2 & 5.2 & 3.9 & 2.6 & 16.3 & 18.5 & 17.5 & 21.8 & 3.2  & 6.0  \\ \hline
G & 3.6 & 5.0 & 3.2 & 2.4 & 16.3 & 19.1 & 17.6 & 23.1 & 1.9  & 5.3  \\ \hline
H & 3.2 & 4.1 & 2.1 & 1.3 & 15.1 & 17.5 & 16.6 & 22.2 & 1.6  & 5.5  \\ \hline
I & 1.8 & 2.7 & 0.9 & 0.6 & 13.3 & 15.8 & 14.7 & 23.0 & 1.0  & 4.2  \\ \hline
J & 2.0 & 3.0 & 1.3 & 0.8 & 13.2 & 15.2 & 14.6 & 23.9 & 1.1  & 4.5  \\ \hline
K & 1.2 & 2.1 & 0.2 & -0.4 & 12.2 & 15.2 & 13.5 & 21.7 & -0.2 & 4.0  \\ \hline
L & 0.7 & 1.7 & -0.1 & -0.4 & 12.2 & 15.1 & 13.5 & 22.7 & 0.1  & 4.2  \\ \hline
\end{tabular}
\caption{The barrier heights in kcal/mol.}
\label{tab:table1}
\end{table}

\begin{table}[H]
\centering
\begin{tabular}{|c|c|c|c|c|c|c|c|c|c|c|}
\hline
\backslashbox{Alkane}{Radical}   & 1    & 2    & 3    & 4    & 5    & 6    & 7    & 8    & 9    & 10   \\ \hline
A & -0.8 & -0.5 & -2.1 & -2.9 & 14.9 & 19.2 & 16.5 & 22.2 & -1.8 & -0.4 \\ \hline
B & 0.2  & 0.5  & -1.0 & -1.9 & 16.0 & 20.2 & 17.6 & 23.3 & -0.8 & 0.7  \\ \hline
C & 0.5 & 1.1 & -0.7 & -1.6 & 16.3 & 20.5 & 17.9 & 23.6 & -0.5 & 1.0  \\ \hline
D & 0.8  & 1.1 & -0.5 & -1.3 & 16.5 & 20.8 & 18.1 & 23.8 & -0.2 & 1.2  \\ \hline
E & -4.0 & -3.7 & -5.3 & -6.1 & 11.7 & 16.0 & 13.3 & 19.0 & -5.0 & -3.6 \\ \hline
F & -3.1 & -2.8 & -4.4 & -5.2 & 12.6 & 16.9 & 14.2 & 19.9 & -4.1 & -2.6 \\ \hline
G & -3.4 & -2.8 & -4.7 & -5.5 & 12.3 & 16.6 & 13.9 & 19.6 & -4.4 & -3.0 \\ \hline
H & -3.1 & -2.8 & -4.3 & -5.2 & 12.7 & 16.9 & 14.3 & 19.9 & -4.1 & -2.6 \\ \hline
I & -6.2 & -5.9 & -7.5 & -8.3 & 9.5  & 13.8 & 11.2 & 16.8 &  -7.2 & -5.7 \\ \hline
J & -5.8 & -5.5 & -7.1 & -7.9 & 9.9  & 14.2 & 11.5 & 17.2 &  -6.8 & -5.3 \\ \hline
K & -6.4 & -5.8 & -7.7 & -8.6 & 9.3  & 13.5 & 10.9 & 16.6 &  -7.5 & -6.0 \\ \hline
L & -6.0 & -5.8 & -7.3 & -8.2 & 9.7  & 14.0 & 11.3 & 17.0 &  -7.1 & -5.6 \\ \hline
\end{tabular}
\caption{The reaction enthalpies in kcal/mol.}
\label{tab:table2}
\end{table}

\section{Discussion}

\subsection{The Evans-Polanyi relationship for sterically unhindered reactions}

To assess the accuracy of the Evans-Polanyi relationship for sterically unhindered reactions, we compare the reactivities of unhindered primary (\textbf{A}), secondary (\textbf{E}), and tertiary (\textbf{I}) C–H bonds in alkanes for three unhindered alkoxy (\textbf{1}), cycloalkoxy (\textbf{9}), and phenoxy (\textbf{5}) radicals.  Taking differences of the Evans-Polanyi relationship for two reactions with the same $E_0$ and $\alpha$ parameters yields:

\begin{equation} \label{eq:1b}
\Delta \Delta H^{\ddag} = \alpha \Delta \Delta H
\end{equation}

If, further, the two HAT reactions being compared involve the same oxy radical, then $\Delta \Delta H$ depends only on the alkane (it is the difference in bond dissociation enthalpies of the C–H bonds in the alkanes of the two reactions), whereas $\Delta \Delta H^{\ddag}$ is different for different oxy radicals.

Table III shows the changes in reaction enthalpies and barrier heights in moving from sterically unhindered primary (\textbf{A}) to secondary (\textbf{E}) to tertiary (\textbf{I}) alkane C–H bonds.  While the relative reaction enthalpies are independent of the specific oxy radical, the relative barrier heights are shown for sterically unhindered alkoxy (\textbf{1}), cycloalkoxy (\textbf{9}), and phenoxy (\textbf{5}) radicals.

\begin{table}[H]
\centering
\begin{tabular}{|c|c|c|c|}
\hline
              & EtH ($1^{\circ}$) (\textbf{A})     &  n-PrH ($2^{\circ}$) (\textbf{E})      &  i-BuH ($3^{\circ}$) (\textbf{I})      \\ \hline
$\Delta \Delta H$ & 0 & -3.2 & -5.4 \\ \hline
Methoxy (\textbf{1}) $\Delta \Delta H^{\ddag}$ & 0 & -2.8 & -5.1 \\ \hline
Cyclohexoxy (\textbf{9}) $\Delta \Delta H^{\ddag}$ & 0 & -3.0 & -4.7 \\ \hline
Phenoxy (\textbf{5}) $\Delta \Delta H^{\ddag}$ & 0 & -3.8 & -6.2 \\ \hline
\end{tabular}
\caption{Relative reaction enthalpies and relative barrier heights in moving from $1^{\circ}$ to $2^{\circ}$ to $3^{\circ}$ alkanes for hydrogen abstraction in kcal/mol.}
\label{tab:table 3}
\end{table}

As shown in the first row of the table, the relative reaction enthalpies $\Delta \Delta H$ becomes more negative as the alkane C–H bond becomes more substituted because the product radical contains more alkyl groups which can stabilize the radical via hyperconjugation.  Surprisingly, with the exception of the reaction between phenoxy radical \textbf{5} and isobutane \textbf{I}, the relative barrier heights $\Delta \Delta H^{\ddag}$ track the relative reaction enthalpies $\Delta \Delta H$ quite closely for all three unhindered alkoxy radicals.  This implies that $\alpha$ in the Evans-Polanyi relationship is approximately 1. The small (-0.8 kcal/mol) negative deviation in $\Delta \Delta H^{\ddag}$ for the reaction between the phenoxy radical and isobutane indicates  stabilization in the transition state more than the product, likely due to dispersion interaction.

\subsection{Electronic effects of phenoxy radicals in HAT reactions}

In order to study the steric effects of phenoxy radicals on the rates of HAT reactions, we first had to determine the electronic effects of alkyl groups.  Thus, we compared the reactivities of unhindered phenoxy (\textbf{5}) and p-tert-butylphenoxy (\textbf{7}) radicals for a series of increasingly substituted alkanes, namely ethane (\textbf{A}), propane (\textbf{B}), isobutane (\textbf{C}), and neopentane (\textbf{D}).  Since, this time, we are making comparisons between different oxy radicals with individual alkanes, $\Delta \Delta H$ depends only on the oxy radicals (it is the difference in bond dissociation enthalpies of the O–H bonds in the phenols of the two reactions), whereas $\Delta \Delta H^{\ddag}$ is different for different alkane substrates.  The results are shown in Table IV below:

\begin{table}[H]
\centering
\begin{tabular}{|c|c|c|}
\hline
              & PhO (\textbf{5})     &  \textit{p}-t-BuPhO (\textbf{7}) \\ \hline
 $\Delta \Delta H$ & 0 & 1.6 \\ \hline
ethane (\textbf{A}) $\Delta \Delta H^{\ddag}$ & 0 & 1.1 \\ \hline
propane (\textbf{B}) $\Delta \Delta H^{\ddag}$ & 0 & 1.5 \\ \hline
isobutane (\textbf{C}) $\Delta \Delta H^{\ddag}$ & 0 & 1.3  \\ \hline
neopentane (\textbf{D}) $\Delta \Delta H^{\ddag}$ & 0 & 1.1 \\ \hline
\end{tabular}
\caption{Relative reaction enthalpies and relative barrier heights of hydrogen abstraction for phenoxy and \textit{p}-t-Bu-phenoxy radicals (in kcal/mol).}
\label{tab:table 4}
\end{table}

Because \textit{para}-substitution in \textbf{7} does not create any additional steric interactions in the HAT transition state or in the newly formed phenol O–H bond in the product, comparing the HAT reactivity of phenoxy radicals \textbf{5} and \textbf{7} isolates the electronic effects.  In particular, the \textit{p}-tBu group in \textbf{7} simply stabilizes the phenoxy radical starting material via hyperconjugation, thus increasing the relative enthalpies of both the transition state and the product.  In particular, the reaction enthalpy ($\Delta H$) increases by 1.6 kcal/mol, while barrier heights increase by about 1.3 kcal/mol on average, consistent with an Evans-Polanyi $\alpha$ of approximately 0.8.

\subsection{Steric effects of phenoxy radicals in HAT reactions}

Having considered the electronic effect of a p-tert-butyl, we investigated steric effects of phenoxy radicals on the rates of HAT reactions by considering ortho-substituted phenoxy radicals. In particular, we compare the reactivities of increasingly hindered phenoxy (\textbf{5}), 2,6-dimethylphenoxy (\textbf{6}), and 2,6-di-tert-butylphenoxy (\textbf{8}) with ethane (\textbf{A}), propane (\textbf{B}), isobutane (\textbf{C}), and neopentane (\textbf{D}) are summarized in Table V.  As in the previous section, $\Delta \Delta H$ is the same for any alkane substrate while $\Delta \Delta H^{\ddag}$ is not.

\begin{table}[H]
\centering
\begin{tabular}{|c|c|c|c|}
\hline
              & PhO (\textbf{5})     &  di-MePhO (\textbf{6}) &  di-t-BuPhO (\textbf{8}) \\ \hline
$\Delta \Delta H$ & 0 & 4.3 & 7.3 \\ \hline              
ethane (\textbf{A}) $\Delta \Delta H^{\ddag}$ & 0 & 2.0 & 6.8 \\ \hline
propane (\textbf{B}) $\Delta \Delta H^{\ddag}$ & 0 & 2.1 & 6.2  \\ \hline
isobutane (\textbf{C}) $\Delta \Delta H^{\ddag}$ & 0 & 3.7 & 6.2 \\ \hline
neopentane (\textbf{D}) $\Delta \Delta H^{\ddag}$ & 0 & 2.8 & 5.6 \\ \hline
\end{tabular}
\caption{Relative reaction enthalpies and relative barrier heights for hydrogen abstraction for phenoxy and ortho-substituted phenoxy radicals (kcal/mol).}
\label{tab:table 5}
\end{table}

As expected, for all four alkanes considered, the barrier heights $ \Delta H^{\ddag}$ increase as the \textit{ortho}-substituents become bulkier from H  (\textbf{5}) to Me (\textbf{6}) to t-Bu (\textbf{7}), due to increasing steric repulsion in the transition states between these \textit{ortho}-substituents and atoms of the alkane.  However, perhaps more surprisingly, there is a corresponding increase in the reaction enthalpies even though $\Delta H$ reflects the difference in enthalpy between \textit{separated} products and \textit{separated} starting materials, meaning that no interactions between the \textit{ortho}-substituents and the atoms of the alkane is possible. We also note a negative deviation in $ \Delta \Delta H^{\ddag}$ for the reactions of neopentane \textbf{D} with both \textit{ortho}-substituted phenoxy radicals, likely due to dispersion (see later).

To understand why bulky ortho-substituents lead to higher reaction enthalpies, we begin by noting that the same electronic stabilization described in the previous section for \textit{para}-substituted phenoxy radical \textbf{7} also exists for \textit{ortho}-substituted \textbf{6} and \textbf{8}.  Given that \textbf{6} and  \textbf{8} each have two alkyl substituents, we can estimate this electronic stabilization as 2 × 1.6 kcal/mol = 3.2 kcal/mol.  However, this still leaves approximately 1.1 kcal/mol in  \textbf{6} and 5.1 kcal/mol in  \textbf{8} which cannot be explained by electronic effects.

To explain these remaining contributions to $\Delta \Delta H$, we identify a  steric effect present in the alcohol product. In particular, as the \textit{ortho}-substituents become bulkier, i.e., from H to Me to tBu, the hydrogen on the phenol experiences more steric repulsion, raising the enthalpy of the alcohol product.  Figure 4 compares the shortest H–H distance between the phenol hydrogen and the \textit{ortho}-substituents in the phenol products of \textbf{5}, \textbf{6}, and \textbf{8}.

\begin{figure}[H]
\centering
\includegraphics[width=1.3\linewidth]{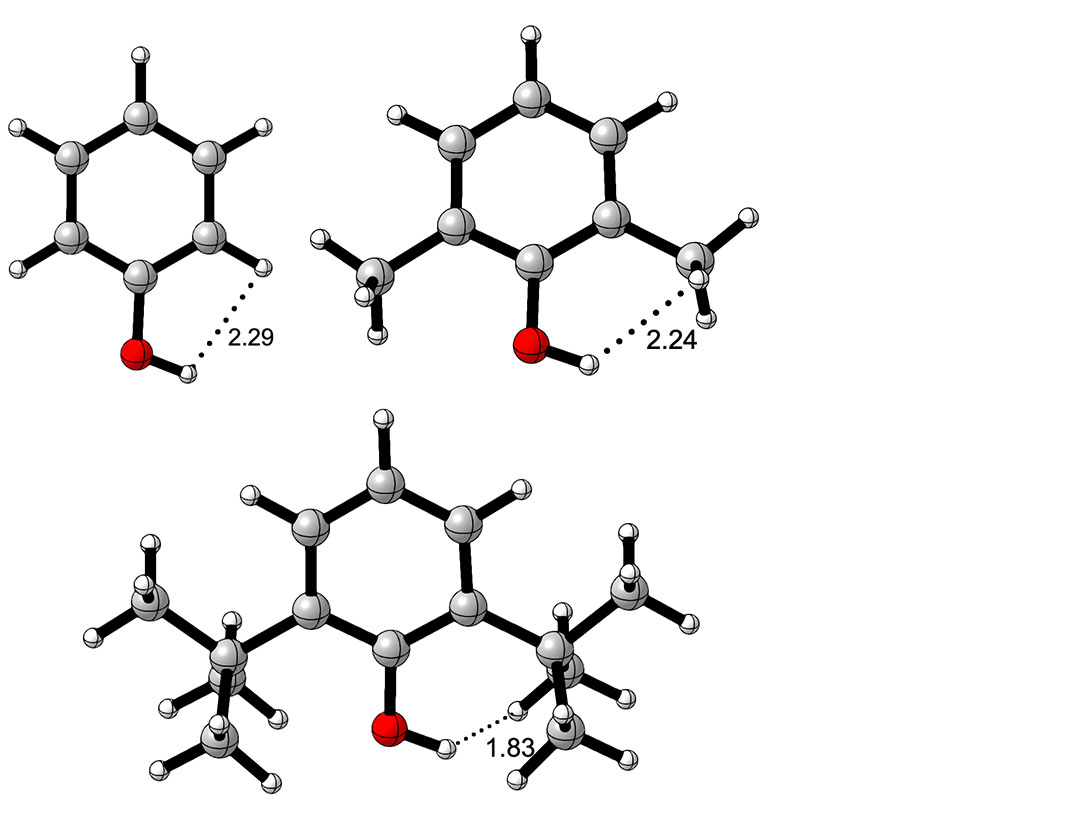}
\caption{Comparing the shortest H–H distance (in Å) between the phenol hydrogen and the \textit{ortho}-substituents in phenol (top), 2,6-dimethylphenol (center), and 2,6-di-\textit{tert}-butylphenol (bottom).  For comparison, twice the van der Waals radius of hydrogen is 2.4 Å.}
\end{figure}

As the figure shows, the H–H bond distance decreases from 2.29 Å in phenol to 2.24 Å in 2,6-dimethylphenol to 1.83 Å in 2,6-di-\textit{tert}-butylphenol.  Given that twice the van der Waals radius of hydrogen is 2.4 Å, this represents steric repulsion between the H of the newly-formed O–H bond and the \textit{ortho}-substituents in the product phenols.  In this way, bulky \textit{ortho}-substituents destabilize product phenols relative to the starting phenoxy radicals and thus increase the HAT reaction enthalpy $\Delta H$.  \textit{Ortho}-substituents destabilize \textit{both} the transition states \textit{and} the phenol products relative to the starting materials.

This means the Evans-Polanyi relationship between $\Delta H^{\ddag}$ and
$\Delta H$ also holds when steric repulsions are present. Quantitatively, \textit{ortho}-methyl groups increase barrier heights by an average of 2.7 kcal/mol and reaction enthalpies by 4.3 kcal/mol, while \textit{ortho}-\textit{tert}-butyl groups increase barrier heights by an average of 6.2 kcal/mol and reaction enthalpies by 7.3 kcal/mol, leading to an Evans-Polanyi fitting parameter $\alpha$ in the range of 0.6 to 0.85, not far from the estimated value of 0.8 in the absence of steric bulk.

Moving beyond phenoxy radicals, we observe a similar concomitant increase in HAT barrier heights and reaction enthalpies when comparing the cyclohexoxy radical (\textbf{9}) with the sterically-hindered 2,2,6,6-tetramethylcyclohexoy radical (\textbf{10}), as shown in Table VI below. Now the steric effect destabilizes the TS twice as much as the product.

\begin{table}[H]
\centering
\begin{tabular}{|c|c|c|}
\hline
              & CyO (\textbf{9})     &  2,2,6,6-tetra-MeCyO  (\textbf{10}) \\ \hline
$\Delta \Delta H$ & 0 & 1.5  \\ \hline              
ethane (\textbf{A}) $\Delta \Delta H^{\ddag}$ & 0 & 3.0 \\ \hline
propane (\textbf{B}) $\Delta \Delta H^{\ddag}$ & 0 & 2.8  \\ \hline
isobutane (\textbf{C}) $\Delta \Delta H^{\ddag}$ & 0 & 2.8 \\ \hline
neopentane (\textbf{D}) $\Delta \Delta H^{\ddag}$ & 0 & 3.0 \\ \hline
\end{tabular}
\caption{Relative reaction enthalpies and relative barrier heights for hydrogen abstraction for cyclohexoxy and 2,2,6,6-tetramethylcyclohexoxy radicals (kcal/mol).}
\label{tab:table 6}
\end{table}

\subsection{Steric attraction in HAT reactions of alkoxy radicals and alkanes}

Finally, we consider the role of steric attraction, or dispersion interactions \cite{Bondi}, in the barrier heights of HAT reactions between alkoxy radicals and alkanes.  Among the alkoxy radicals in our study, the bulkiest is \textit{tert}-butoxy radical \textbf{4} while the least bulky is methoxy radical \textbf{1}. With these two alkoxy radicals, we can compare the most hindered and least hindered reactions in our study for primary, secondary, and tertiary alkane C–H bonds.  For primary alkane C–H bonds, the least hindered HAT reaction is between methoxy radical \textbf{4} and ethane \textbf{A}, while the most hindered reaction is between \textit{tert}-butoxy radical \textbf{4} with neopentane \textbf{D}.  Similarly, for secondary alkane C–H bonds, the least hindered reaction is between \textbf{1} and \textbf{E} and the most hindered reaction is between \textbf{4} and \textbf{H}; for tertiary alkane C–H bonds, the least hindered reaction is between \textbf{1} and \textbf{I} and the most hindered reaction is between \textbf{4} and \textbf{L}.  Table VII compares the relative barrier heights and relative reaction enthalpies between these least hindered and most hindered HAT reactions for primary, secondary, and tertiary C–H bonds.

\begin{table}[H]
\centering
\begin{tabular}{|c|c|c|c|}
\hline
            & \makecell{Primary C–H \\ (A,1) to (D,4)} &  \makecell{Secondary C–H \\ (E,1) to (H,4)} & \makecell{Tertiary C–H \\ (I,1) to (L,4) }   \\ \hline
$\Delta \Delta H$ & -0.5 & -1.2 & -2.0 \\ \hline
$\Delta \Delta H^{\ddag}$ & -2.5 & -2.8 & -2.2 \\ \hline
\end{tabular}

\caption{Reaction enthalpies and barrier heights for least hindered and most hindered HAT reactions between alkoxy radicals and primary, secondary, and tertiary alkane C–H bonds in kcal/mol.
}
\label{tab:table 7}
\end{table}

As the table shows, for primary, secondary, and tertiary C–H bonds, both the reaction enthalpy and the barrier heights decrease when the reactants go from unhindered to bulky.  Presumably, the decrease in reaction enthalpy $\Delta H$ is caused by increased hyperconjugative stabilization of the product alkyl radical when the alkane is bulkier.  However, the even greater decrease in the barrier heights $\Delta H^{\ddag}$ suggests the presence of stabilizing dispersion interactions between the two bulky reactants in the transition state. The origins of these dispersion factors are shown graphically in Figure 5. Here, the transition state for the reaction of \textbf{4} with \textbf{D} is shown, with the shortest distances between hydrogens on the two reactants labeled. These distances reveal that the hydrogens are close enough for steric attraction.

\begin{figure}[H]
\centering
\includegraphics[width=0.75\linewidth]{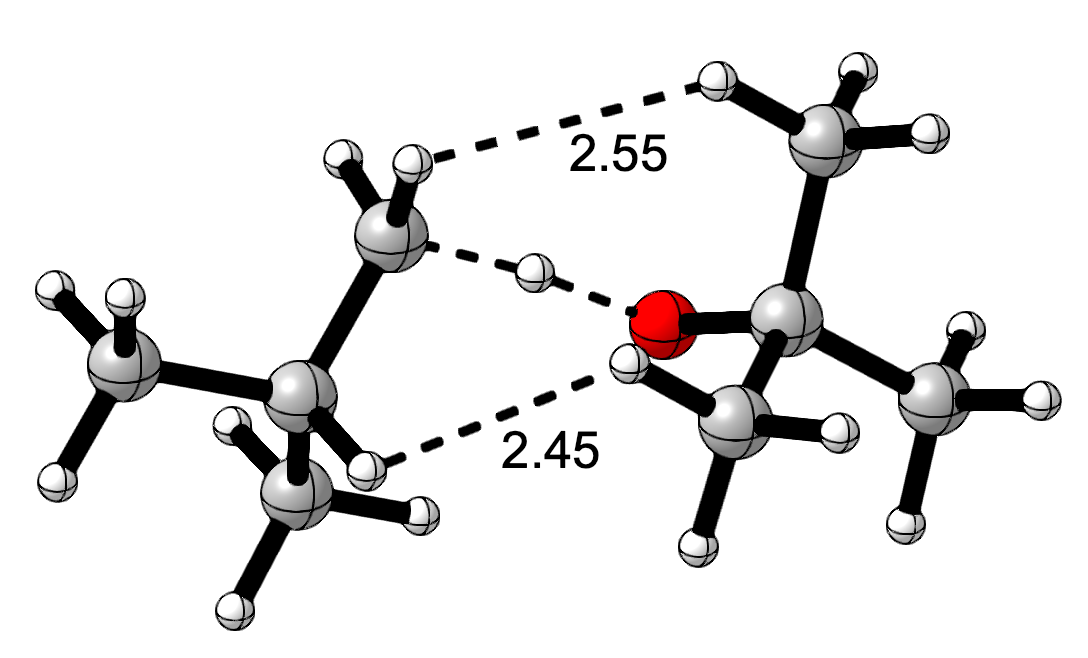}
\caption{Transition state for the reaction of 4 with D, with distances between the nearest hydrogens on the two reactants labeled.}
\end{figure}

To summarize, as the reactants in HAT reactions become more bulky, the decrease in reaction enthalpy is primarily an electronic effect caused by larger hyperconjugative stabilization of the product alkyl radical, while the additional decrease in barrier heights is primarily a steric attraction effect caused by stronger dispersion interactions that stabilize the transition state.

\section{Conclusion}

In this study, the role of steric effects in the HAT process is explored by analyzing 120 HAT reactions. Reaction enthalpies and barrier heights are controlled by three different effects: steric repulsion, electronic stabilization, and steric attraction or dispersion effects.  Steric repulsion effects appear not only in the transition states but also in the products.

\section*{Acknowledgements}
We thank Huiling Shao for helpful discussions.  K. N. H. acknowledges the support of the National Science Foundation under award number CHE-2153972.  Computational resources were provided by the University of Cambridge.



\end{document}